# Structural and dielectric properties of glasses in the system $TeO_2$-$CaCu_3Ti_4O_{12}$.

P. Thomas[a] and K.B.R. Varma.[b*]

[a] Dielectric Materials Division, Central Power Research Institute, Bangalore:560080, India

[b] Materials Research Centre, Indian Institute of Science, Bangalore: 560012, India

Abstract

The glasses in the system $(100-x)TeO_2$-$xCaCu_3Ti_4O_{12}$, ($x$=0.25 to 3 mol %) were fabricated. The color varied from olive green to brown as the $CaCu_3Ti_4O_{12}$ (CCTO) content increased in $TeO_2$ matrix. The X-ray powder diffraction and differential scanning calorimetric analyses that were carried out on the as-quenched samples confirmed their amorphous and glassy nature respectively. The dielectric constant and loss in the 100 Hz-1MHz frequency range were monitored as a function of temperature (50-400°C). The dielectric constant ($\varepsilon_r'$) and the loss (D) increased as the CCTO content increased in $TeO_2$ at all the frequencies and temperatures under investigation. Further, the $\varepsilon_r'$ and D were found to be frequency independent in the 50-200°C temperature range. The value obtained for the loss at 1MHz was 0.0019 which was typical of low loss materials, and exhibited near constant loss (NCL) in the 100Hz-1MHz frequency range. The electrical relaxation was rationalized using the electric modulus formalism. These glasses may be of considerable interest as substrates for high frequency circuit elements in conventional semiconductor industries owing to their high thermal stability.

***Keywords:*** Dielectric; $TeO_2$; $CaCu_3Ti_4O_{12}$; Glass;

* Corresponding author : Tel. +91-80-2293-2914;  Fax: +91-80-2360-0683.
E-mail : kbrvarma@mrc.iisc.ernet.in (K.B.R.Varma)

## I. Introduction

The dielectrics based on glasses have high dielectric constants, associated with low dissipation factor [1-5] that change very little with frequency and temperature. They do not age (as there are no grain\boundaries and pores) as there is no subsequent chemical reaction taking place, and can be operated at high temperatures without any damage or performance shortfall. They also find applications in temperature sensors, oscillators, filters, and microwave windows.[6-8] Tellurium oxide ($TeO_2$) based glasses are known to possess large refractive indices, relatively high dielectric constants and third order nonlinear optical (NLO) susceptibilities.[9-12] These are also known to have high densities, low glass transition temperatures,[13,14] low melting temperatures and possess large infrared transmission window.[11,15,16] Since $TeO_2$ is a good network former and important material for electrical and optical applications,[17-19] many compositions of tellurite glasses containing non-transition [20] and transition metal oxides were fabricated.[21]

Tellurite based glasses exhibit high dielectric constants and electrical conductivity as compared to the other glass systems owing to the unshared pair of electrons of the $TeO_4$ group that do not take part in the bonding.[22] The dielectric properties of tellurite based glasses vary depending on the type and level of the network modifiers present in the glass structure and their field strength.[23-25] There is a need for glasses with improved dielectric properties (high dielectric constants associated with low loss) that can be exploited in conventional semiconductor processing, apart from retaining their glassy characteristics.[25] The $CaCu_3Ti_4O_{12}$ (CCTO) ceramic which has centrosymmetric bcc structure (space group Im3, lattice parameter $a \approx 7.391$, and $Z=2$), has gained considerable attention due to its unusual high dielectric constant ($\varepsilon_r' \approx 10^{4-5}$) which is nearly independent of frequency (upto 10 MHz) and low thermal coefficient of permittivity (TCK) over a wide range (100-600K) of temperatures.[27-29] Therefore, it was worth investigating into the glass forming ability and physical properties including dielectric properties of TeO$_2$-CCTO system. This article reports the details concerning the fabrication, structural, and dielectric properties of CaCu$_3$Ti$_4$O$_{12}$ containing TeO$_2$ glasses at various frequencies (100Hz-1MHz).

## II. Experimental

Transparent glasses in the system (100-$x$)TeO$_2$-$x$CaCu$_3$Ti$_4$O$_{12}$, where $x$=0.25 to 3, were fabricated via the conventional melt-quenching technique.[3,4] For this, appropriate amounts of CCTO nano powder (pre-synthesized using oxalate precursor route, the details of which could be found elsewhere [30] were mixed with TeO$_2$ and melted at 850°C for 30 min in platinum crucible. Subsequently, the melt was quenched between two stainless steel plates. To prevent the samples from cracking, the steel plates were preheated to 150°C. Optically clear glass plates of about 1-1.5 mm in thickness and 10-15 mm in diameter were obtained by this technique. All these samples were annealed at a temperature which is about 50°C below their glass transition temperature for 5h to get rid of the residual stresses that are likely to exist. X-ray powder diffraction studies were carried out on the as-quenched and heat treated glass plates using an X'PERT-PRO Diffractometer (Philips, Netherlands) using CuK$\alpha_1$ radiation ($\lambda = 0.154056$ $nm$) in a wide range of 2$\theta$ (5° ≤ 2θ ≤ 85°) with 0.0170 step size using the 'Xcelerator' check program. Differential scanning calorimeter (DSC) (model:Diamond DSC,Perkin Elmer) was employed to ascertain the glassy nature of the as-quenched samples. The capacitance measurements on the silver electroded glass plates were carried out as a function of frequency (100Hz–1MHz) using an impedance gain-phase analyzer (HP4194A).

## III. Results and Discussion

Differential Scanning Calorimetric (DSC) studies were carried out on the as-quenched samples in order to monitor the glass transition and crystallization temperatures. The DSC traces that were obtained for the as-quenched glass pieces at a heating rate of $10^o$C/min for the representative compositions $x$=0.25 and 3.0 are depicted in fig.1. These exhibit endotherm and exotherm (peak temperature) around 377 and $502^o$C respectively for the composition corresponding to $x$=0.25. It is observed that both the glass transition (endotherm) and crystallization (exotherm) shifted towards higher temperatures as the CCTO content increased from 0.25 mol% to 3.0 mol% in $TeO_2$. The exothermic peak is very broad suggesting the dominant mechanism to be surface crystallization.

XRD patterns obtained for the as-quenched and heat treated upto $450^o$C/2h for all the compositions (not shown here) did not show any long-range structural order, indicating their amorphous nature. The sample corresponding to the composition $x$=3.0 was heat treated at various temperatures (550-$600^o$C) and interestingly, the X-ray diffraction pattern recorded (fig.2a) for the sample heat treated around $550^o$C/2h had revealed the presence of crystallites. The sample became opaque at this stage of heat treatment. This diffraction pattern is assigned to the crystallization phases of tellurium oxide, calcium titanate and titanium tellurite (Winstanleyite). The XRD pattern obtained

(fig.2 b) for the sample heat treated around 570°C/2h is similar to the one heat treated around 550°C. However, when the sample was heat treated around 600°C/2h, the formation of titanium tellurite (Winstanleyite) is more predominant as revealed by the X-ray diffraction studies (fig.2 c).

The frequency dependent (100Hz-1MHz) characteristics of dielectric constant ($\varepsilon_r'$) and the loss (D) for the as quenched glass samples corresponding to the system (100-$x$)TeO$_2$-xCaCu$_3$Ti$_4$O$_{12}$, where $x$=0.25, 0.5, 1.0, 2.0 and 3.0 are shown in Fig. 3(a&b). It is observed that, there is an increase in $\varepsilon_r'$ as the CCTO content increased from 0.25 mol % to 3 mol % in TeO$_2$. The dielectric constant (fig.3a) for the 0.25 mol% CCTO in TeO$_2$ is 15.2 at 100Hz, which remains almost the same over the whole frequency range ($10^2$-$10^6$Hz) under study. Similarly, it has been observed that there is no appreciable change in the dielectric constant from 100Hz to 1MHz for the 3.0 mol% CCTO in TeO$_2$. For instance, the dielectric constant for the 3.0 mol% CCTO in TeO$_2$ has gone upto 27.1 at 100Hz and decreased to 26.5 at 1MHz which is well within the error involved in making these measurements indicating that the dielectric constant is independent of frequency. The room temperature dielectric loss (@100Hz) value has increased from 0.0079 to 0.021 when the CCTO content is increased from the 0.25 to 3.0 mol % (fig.3b). However, the value at 1MHz is only 0.0019 for all the compositions under investigation which is typical of the most tellurite glasses. In essence, frequency independent (flat) dielectric constant and low loss behaviour was observed in the 100Hz-1MHz frequency range for all the present compositions. The dielectric properties of tellurite glasses are known to depend on the size of the modifier ions in the glass structure, their field strengths and composition of the glass. The CCTO ceramics [31] fabricated with the addition of 1.5% $TeO_2$ by weight, exhibited low dielectric loss (0.09) associated with a dielectric constant over 3300 around 60Hz. This report though dealt with crystalline phases, corroborates our findings in terms of the important role played by dielectric $TeO_2$. The dielectric constant obtained in this work is higher than that of pure $TeO_2$ glass (Ref.9) ($\varepsilon_r' \approx 20$), when the CCTO content in $TeO_2$ is increased beyond 1.0 mol %. Higher dielectric constant values associated with $TeO_2$-CCTO is primarily due to the enhanced dielectric polarization (interfacial polarization) arising from the presence of Cu and/or Ti concentrations.[31] The dielectric loss (fig.3b) also increased with the increase in CCTO content in $TeO_2$.

The frequency dependent dielectric constant ($\varepsilon_r'$) and loss (D) at various temperatures (50-400°C) for the as-quenched samples corresponding to the composition $x$=3.0 under investigation is depicted in Fig.4(a&b). It is observed that the dielectric constant has very weak dependence on frequency upto 150°C. At the other temperatures under study, dielectric constant decreases with increasing frequency and the dispersion increases with the increase in temperature especially in the low frequency regime. This increase in low frequency dispersion of dielectric constant with increase in temperature is attributed to the increase in the electrode/interfacial polarization.[3,4] Similar results were reported in the literature for the other tellurite based glasses.[24] At the other temperatures under study, the dielectric loss behavior is consistent with that of the dielectric constant. A relaxation peak around 150 Hz was encountered in the 350-400°C temperature range which is attributed to the mobile ion polarization combined with electrode polarization.[3,4] At low frequencies, the ions align themselves along the field direction and fully contribute to the total polarization and hence high dielectric constant. As the frequency increases, the polarizability contribution from ionic and orientation sources decreases and finally disappears due to the inertia of the ions. [25] Hence, the ions would not be able to follow the electric field direction and as a result their contribution to the polarization would be less. Therefore, the dielectric constant decreases with increasing frequency at all the temperatures under investigation. The variation in dielectric constant with temperature is sluggish up to about 150°C and subsequently increases rapidly upto 400°C at all the frequencies under study. This rapid increase and an anomaly in $\varepsilon_r'$ are as a consequence of the incidence of the glass-transition around the same temperature. It is well known that the physical properties (heat capacity, viscosity, and thermal expansion coefficient) of a glassy material often change abruptly while passing through the glass-transition and crystallization temperatures. When the viscosity of the glass abruptly decreases in the glass-transition region, the reaction elements such as dipoles and ions easily respond to the external electric field and the dielectric constant increases. The loss spectra obtained did not exhibit any peak in the 1kHz-1MHz frequency range upto 200°C, typical of low loss materials.[32] The dielectric loss (fig.4b) increases beyond 200°C, which is attributed to the increase in conductivity of the glasses.

The temperature dependence of the ac conductivity at different frequencies corresponding to the representative composition x = 3.0 is shown in fig 5a. At high temperature and low frequencies, the curves tend to merge with each other with a constant slope and is attributed to the contribution from the dc conduction, which was reported in the other tellurite glass systems. [24] In order to have further insight into the low dielectric loss behavior of these glasses, its frequency-dependent electrical conductivity at various temperatures is studied. Conductivity, at different frequencies and temperatures, was calculated by using the dielectric data as per the following formula:

$$\sigma_\omega = \omega \varepsilon_o D \varepsilon_r', \tag{1}$$

where $\sigma_\omega$ is the conductivity at an angular frequency $\omega \ (= 2\pi f)$.

The frequency dependence of the conductivity at various temperatures is shown in fig.5b. The phenomenon of the conductivity dispersion in solids is generally analyzed using Jonscher's law [33]

$$\sigma_\omega = \sigma_{dc} + \sigma_{ac} = \sigma_{dc} + Af^n \tag{2}$$

where $\sigma_{dc}$ is the dc conductivity and $A$ is temperature dependent constant and $n$ is the power law exponent which generally varies between 0 to 1. For a wide variety of ionic conductors, it has been observed that at sufficiently low temperatures and/or high frequencies, the conductivity varies linearly or nearly linearly with frequency. Such a behavior is usually known as near constant loss (NCL) behavior which was reported to be a universal feature of many materials.[3,33]

To accommodate the NCL contribution, a linear term of frequency has been introduced in the universal dielectric response (eqn. 3) which describes the ac conductivity of the glasses

$$\sigma_\omega = \sigma_{dc} + Af^n + Bf^{1.0} \tag{3}$$

The conductivity in the 50-250°C temperature range increases with frequency as a function of $Bf^{1.0}$ suggesting the domination of NCL contribution to the ac conductivity in the frequency range under study. The absence of frequency independent conductivity ($\sigma_{dc}$) region within the measured frequency window indicates lack of long-range ionic diffusion in the 50-250°C temperature range. At high temperatures (300-400°C), the conductivity plots could be parameterized using the Eq.3. At these temperatures, the frequency window of NCL behavior decreases with increase in temperature. In addition, random distribution of the ionic charge carriers via activated hopping gives rise to a frequency independent conductivity at lower frequencies. At higher frequencies, the conductivity exhibits dispersion and eventually becomes almost linear. The characteristic frequency is taken as the frequency at which the dispersion deviated from the DC plateau and is also termed as the hopping rate. [34,35]

When the dielectric constant and loss of materials increase exponentially at low frequencies and high temperatures (as in the present case), it is difficult to distinguish the interfacial polarization and conductivity contributions from that of the intrinsic dipolar relaxations. This difficulty is overcome by representing the dielectric data in terms of the electric modulus. Electrical modulus formalism is invoked to rationalize the dielectric response of the present glasses. The use of electric modulus approach also helps in gaining an insight into the bulk response of materials. This would facilitate to circumvent the problems caused by electrical conduction which might mask the dielectric relaxation process.[3] The complex electric modulus ($M^*$) is defined in terms of the complex dielectric constant ($\varepsilon^*$) and is represented as [36]

$$M^* = (\varepsilon^*)^{-1}, \qquad (4)$$

$$M^* = M' + iM'' \frac{\varepsilon_r'}{(\varepsilon_r')^2 + (\varepsilon_r'')^2} + i\frac{\varepsilon_r''}{(\varepsilon_r')^2 + (\varepsilon_r'')^2}, \qquad (5)$$

where, $M'$, $M''$, $\varepsilon'$ and $\varepsilon''$ are the real and imaginary parts of the electric modulus and dielectric constants, respectively. The real and imaginary parts of the electric modulus at different temperatures are calculated using Eq.(5) for the present glasses and are depicted in Fig.6 (a and b). It is observed that at low frequencies (Fig. 6a), $M'$ decreases as the temperature increases and approaches zero. These observations suggest the suppression of electrode polarization. The $M'$ reaches a maximum value corresponding to $M_\infty = (\varepsilon_\infty)^{-1}$ due to relaxation process. [4] It is also noted that the value of $M_\infty$ decreases with the rise in temperature. The imaginary part of the electric modulus (Fig. 6b) is indicative of the energy loss under electric field and the $M''$ peak shifts to higher frequencies with increasing temperature. As reported (Ref.4), this evidently suggests the involvement of temperature dependent relaxation process associated with the present glass samples.

In the present glass system, in order to investigate into transport mechanism, the dc conductivity at different temperatures ($\sigma_{DC}(T)$), was calculated from the electric modulus data. The dc conductivity could be extracted using the following expression. [36]

$$\sigma_{dc}(T) = \frac{\varepsilon_o}{M_\infty(T)^* \tau_m(T)} \left[ \frac{\beta(T)}{\Gamma(1/\beta(T))} \right] \quad (6)$$

where $\varepsilon_o$ is the free space dielectric constant, $M_\infty(T)$ is the reciprocal of high frequency dielectric constant, $\Gamma\left(1/\beta(T)\right)$ is the gamma function value of $1/\beta$ at various temperatures and $\tau_m(T)$ is the temperature dependent relaxation time. Fig.7 shows the dc conductivity data obtained from the above expression [eq.(6)] at various temperatures. The activation energy for the dc conductivity is calculated from the plot of $\ln(\sigma_{dc})$ versus 1000/T (fig.7)

for the present glass. The plot is found to be linear and fitted using the following Arrhenius equation

$$\sigma_{dc}(T) = B \exp\left(\frac{-E_{dc}}{kT}\right) \qquad (7)$$

where B is the pre-exponential factor, $E_{dc}$ is the activation energy for the dc conduction. The activation energy is calculated from the slope of the fitted line and found to be 0.35± 0.03eV. The relaxation time associated with the above process is also determined from the plot of M'' versus frequency.

The activation energy involved in the relaxation process of ions could be obtained from the temperature dependent relaxation time

$$\tau_m = \tau_o \exp\left(\frac{E_R}{kT}\right), \qquad (8)$$

where $E_R$ is the activation energy associated with the relaxation process, $\tau_o$ is the pre-exponential factor, k is the Boltzmann constant, and T is the absolute temperature.

Fig.7 shows a plot between $\ln(\tau_m)$ and 1000/T along with the theoretical fit (solid line) to the above equation [Eq.(8)]. The value that is obtained for $E_R$ is 0.34 ± 0.02 eV, which is in close agreement with that of the activation energy associated with the dc conductivity. It suggests that similar energy barriers are involved in both the relaxation and conduction processes.

## V. Conclusions

The glasses in the system $(100-x)TeO_2-xCaCu_3Ti_4O_{12}$ have been investigated using a number of experimental techniques. The glass transition and crystallization peaks shifted towards higher temperatures as the CCTO content increased from 0.25 mol% to

3.0 mol% in $TeO_2$ matrix. The samples heat treated upto 450$^o$C retained their amorphous characteristics. The dielectric constant for CCTO added $TeO_2$ glasses was higher than that of pure tellurite glasses. Interestingly, the dielectric constant and the loss were frequency independent in the 50-200$^o$C temperature range. The physical properties that were studied on these glasses may shed light on their potential use in high frequency circuit elements in conventional semiconductor industries.

**Acknowledgement**

The management of Central Power Research Institute is acknowledged for the financial support (CPRI Project No. R-DMD-01/1415).

**Figure captions**

Figure.1. DSC thermograms of the as-quenched samples of two different compositions (a) $x=0.25$ and (b) $x=3.0$ mol % CCTO in TeO2 glass.

Figure. 2. XRD patterns for the sample corresponding to the composition $x=3.0$ mol % CCTO containing $TeO_2$ glasses heat treated at (a) 550°C, (b) 570°C and (c) 600°C.

Figure.3. Frequency dependent (a) dielectric constant and (b) dielectric loss for the as quenched samples measured at room temperature (300K) for different CCTO concentrations ($x=0.25$ to 3.0 mol % CCTO in TeO2 glass).

Figure.4. Frequency dependent (a) dielectric constant and (b) dielectric loss measured at various temperatures for the as quenched sample corresponding to the composition $x=3.0$ mol % CCTO containing $TeO_2$ glasses.

Figure 5. Variation of ac conductivity (a) as a function of temperature at various frequencies and (b) as a function of frequency at various temperatures for the as quenched sample corresponding to the composition $x=3.0$ mol % CCTO containing $TeO_2$ glasses.

Figure 6 (a) Real ($M'$) and (b) Imaginary ($M''$) parts of electric modulus as a function of frequency at various temperatures for the as quenched sample corresponding to the composition x=3.0 mol % CCTO containing $TeO_2$ glasses.

Figure 7. Arrhenius plots for the relaxation time and dc conductivity for the as quenched sample corresponding to the composition x=3.0 mol % CCTO containing $TeO_2$ glasses.

Figure.1
Click here to download high resolution image

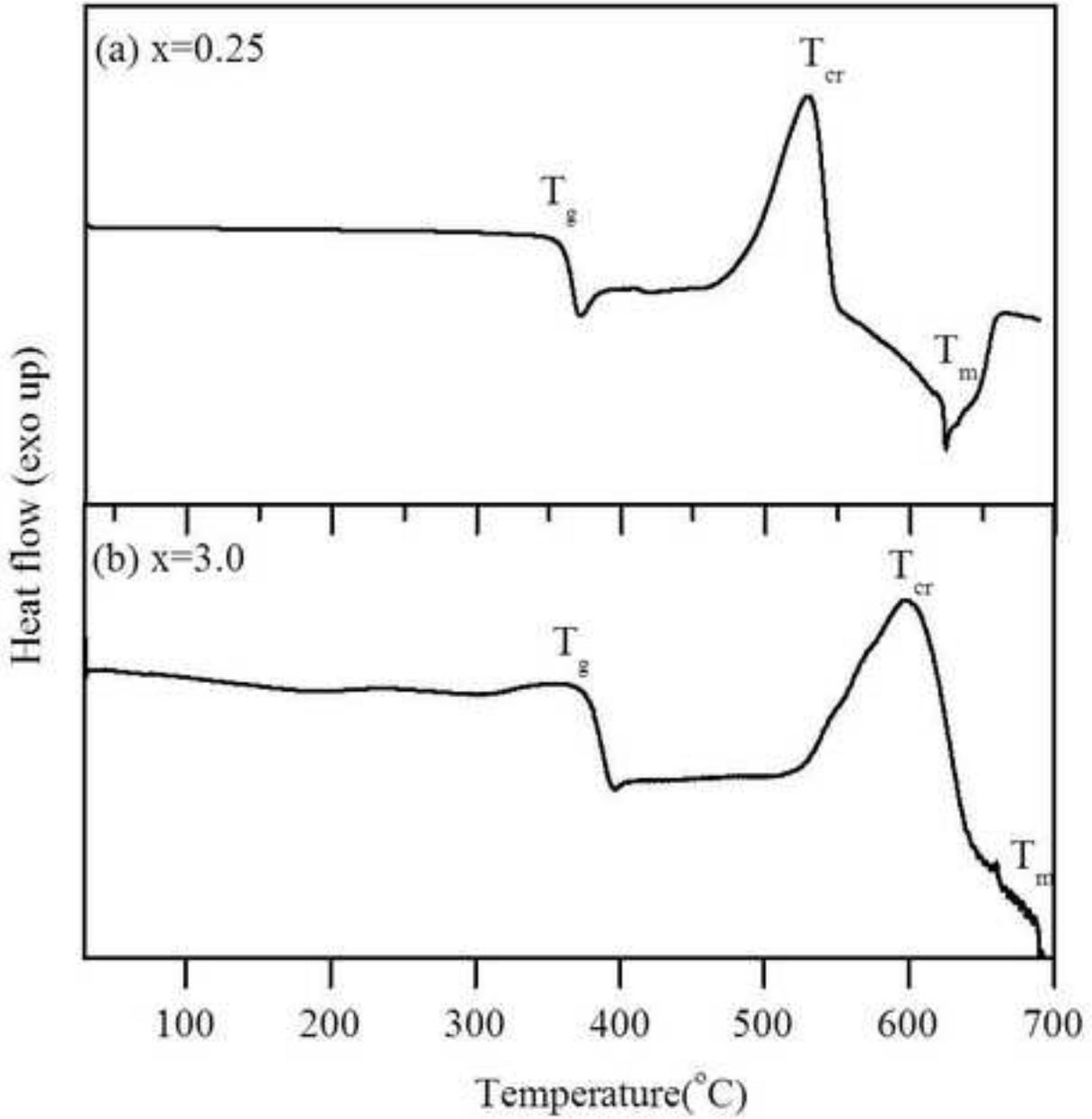

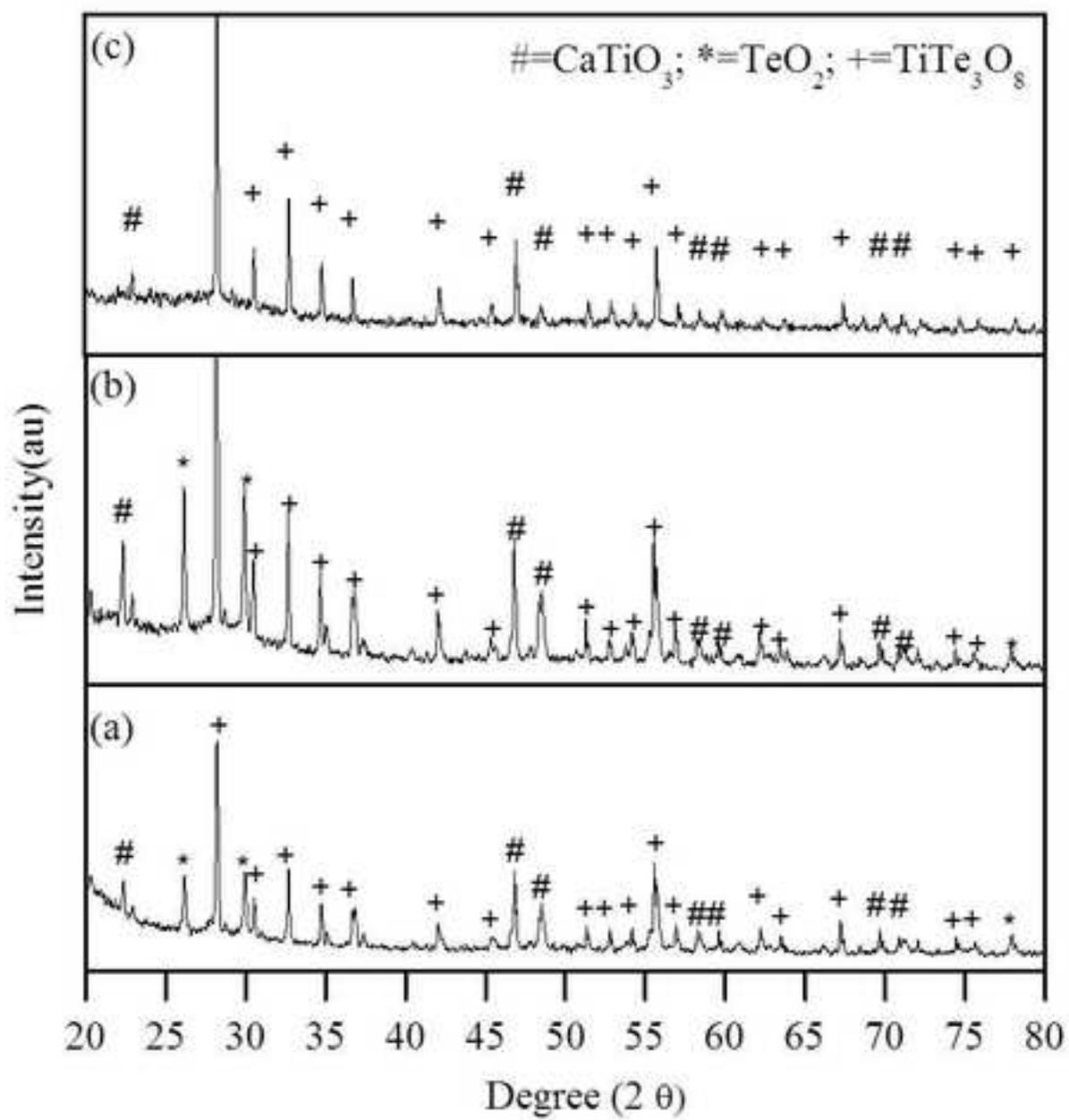

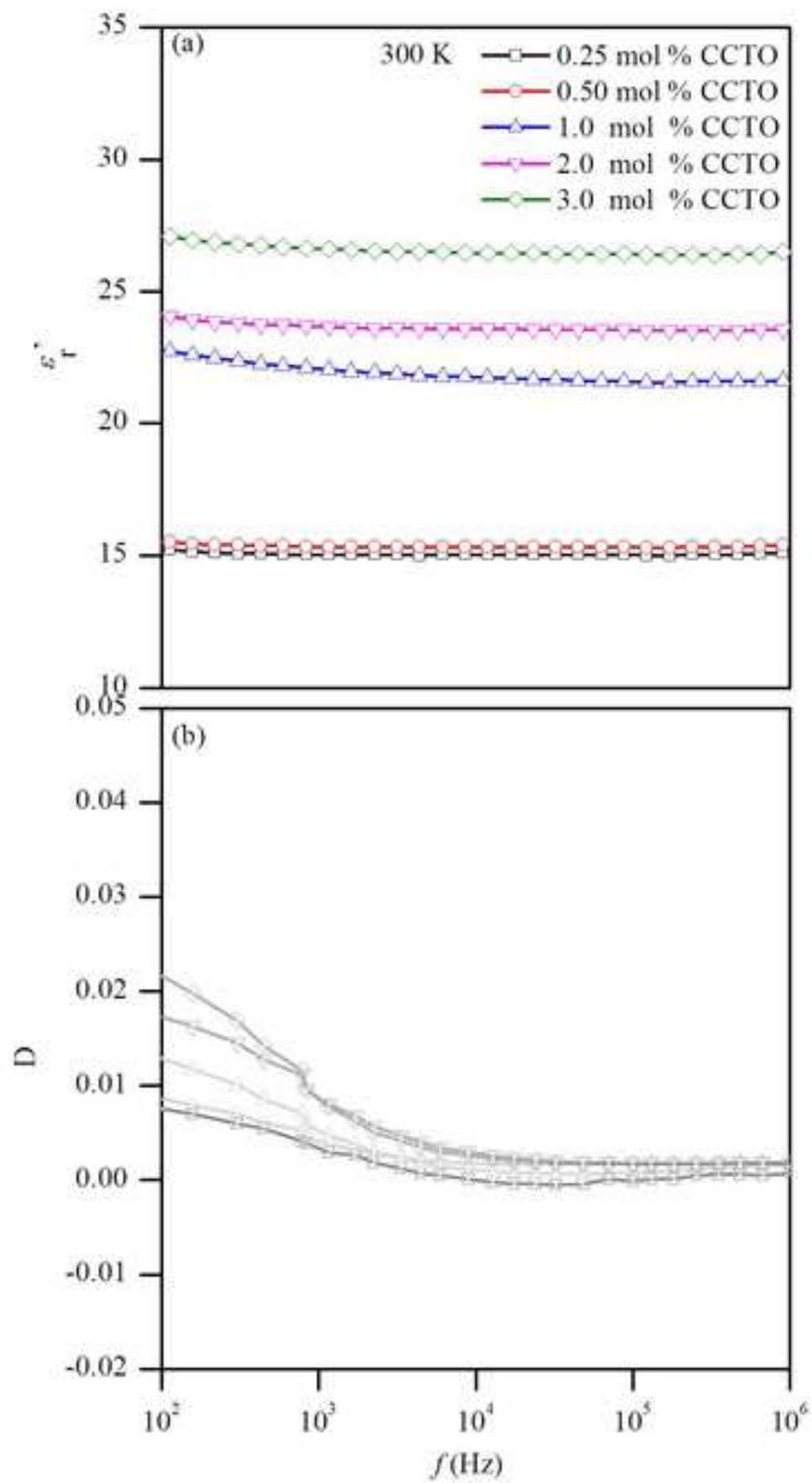

Figure.4
Click here to download high resolution image

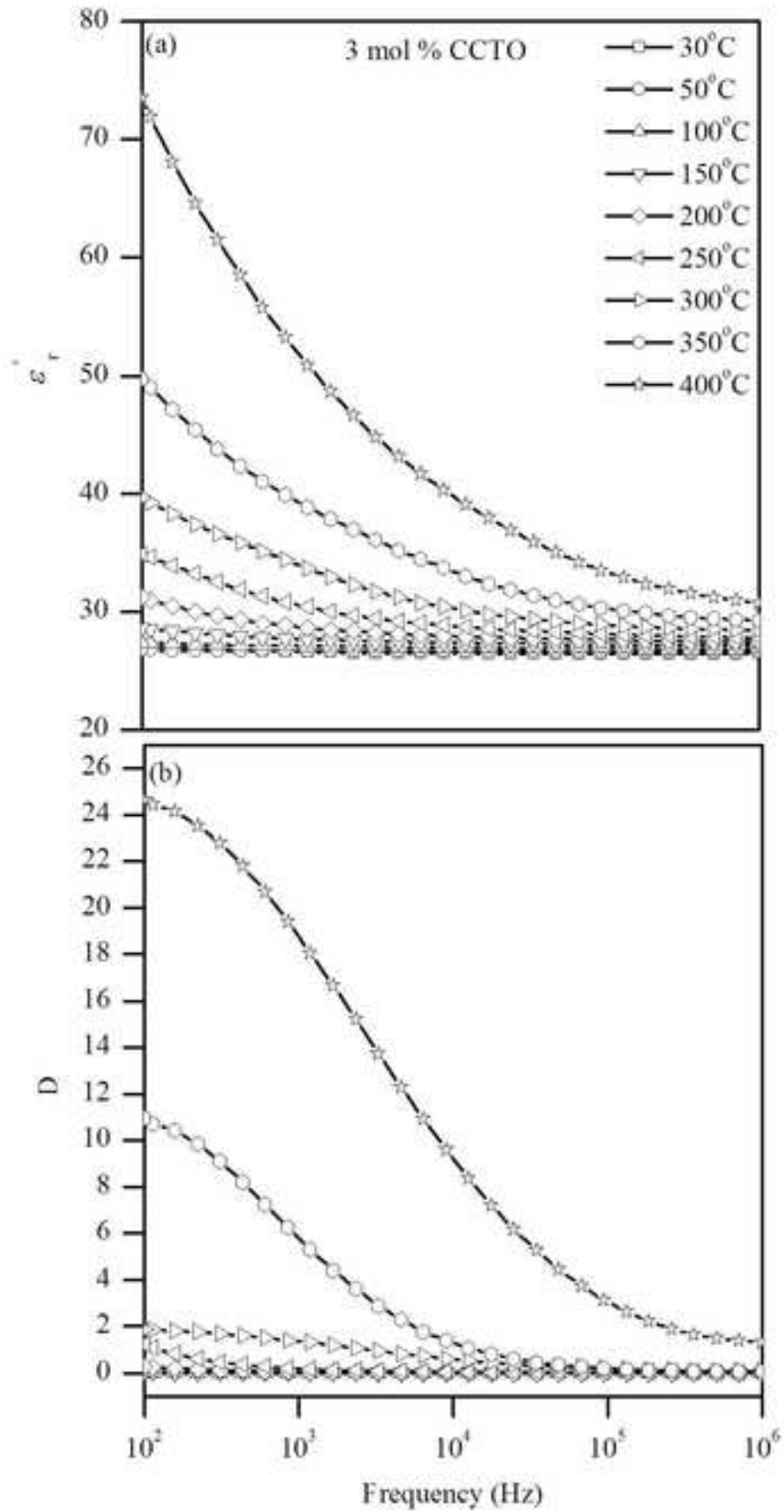

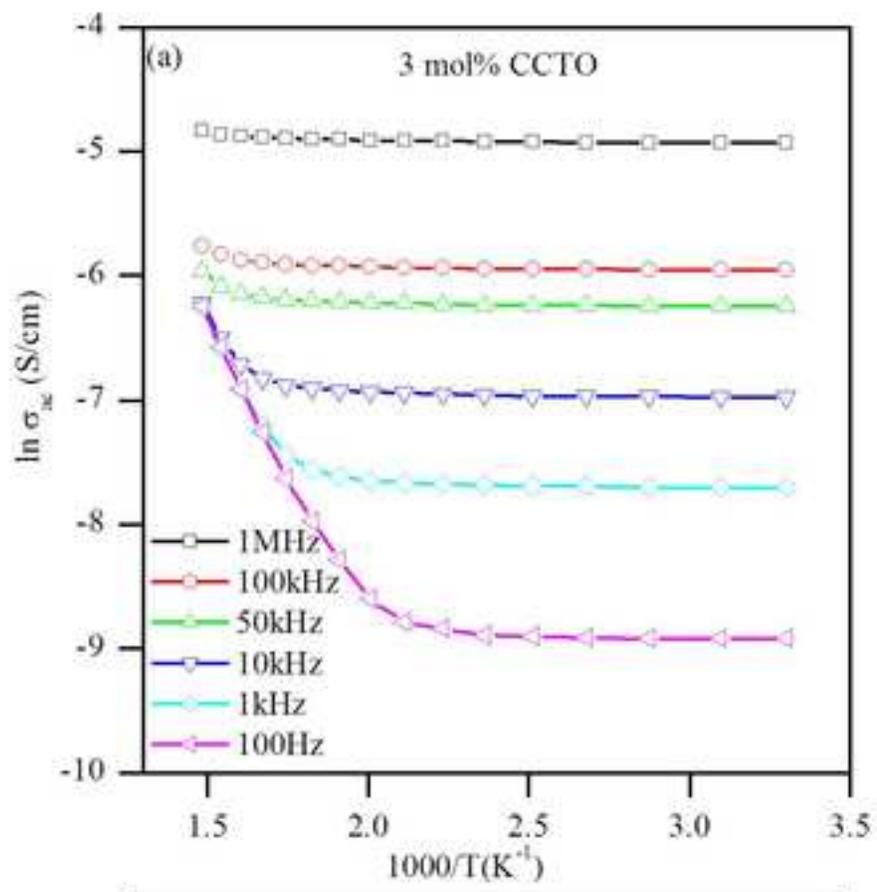

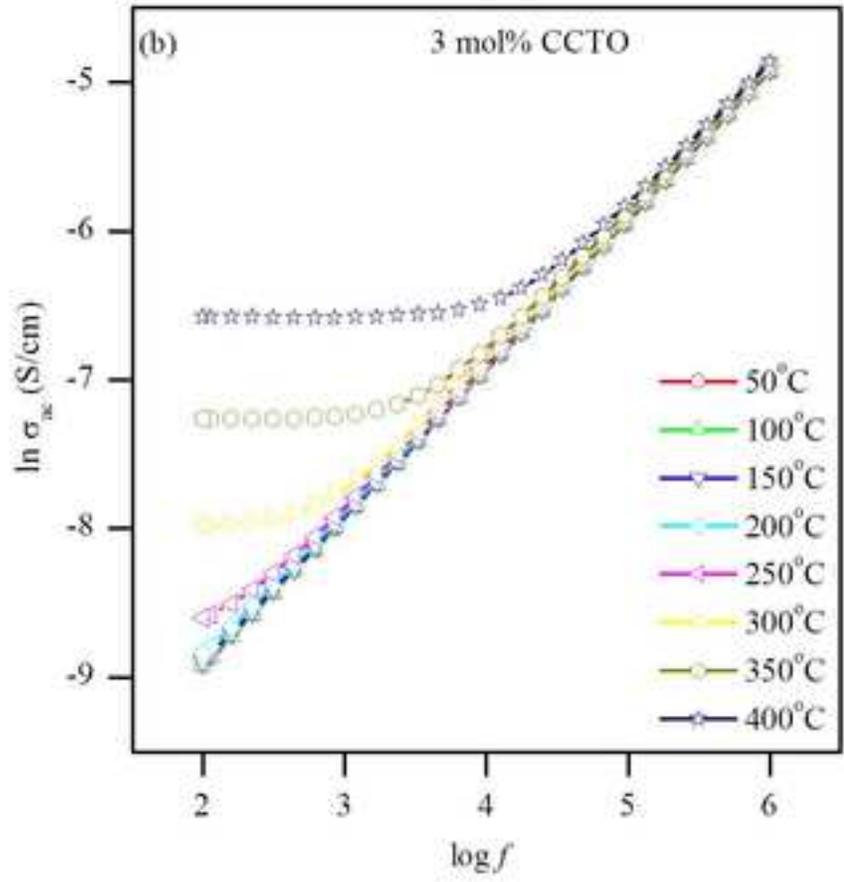

Figure.6
Click here to download high resolution image

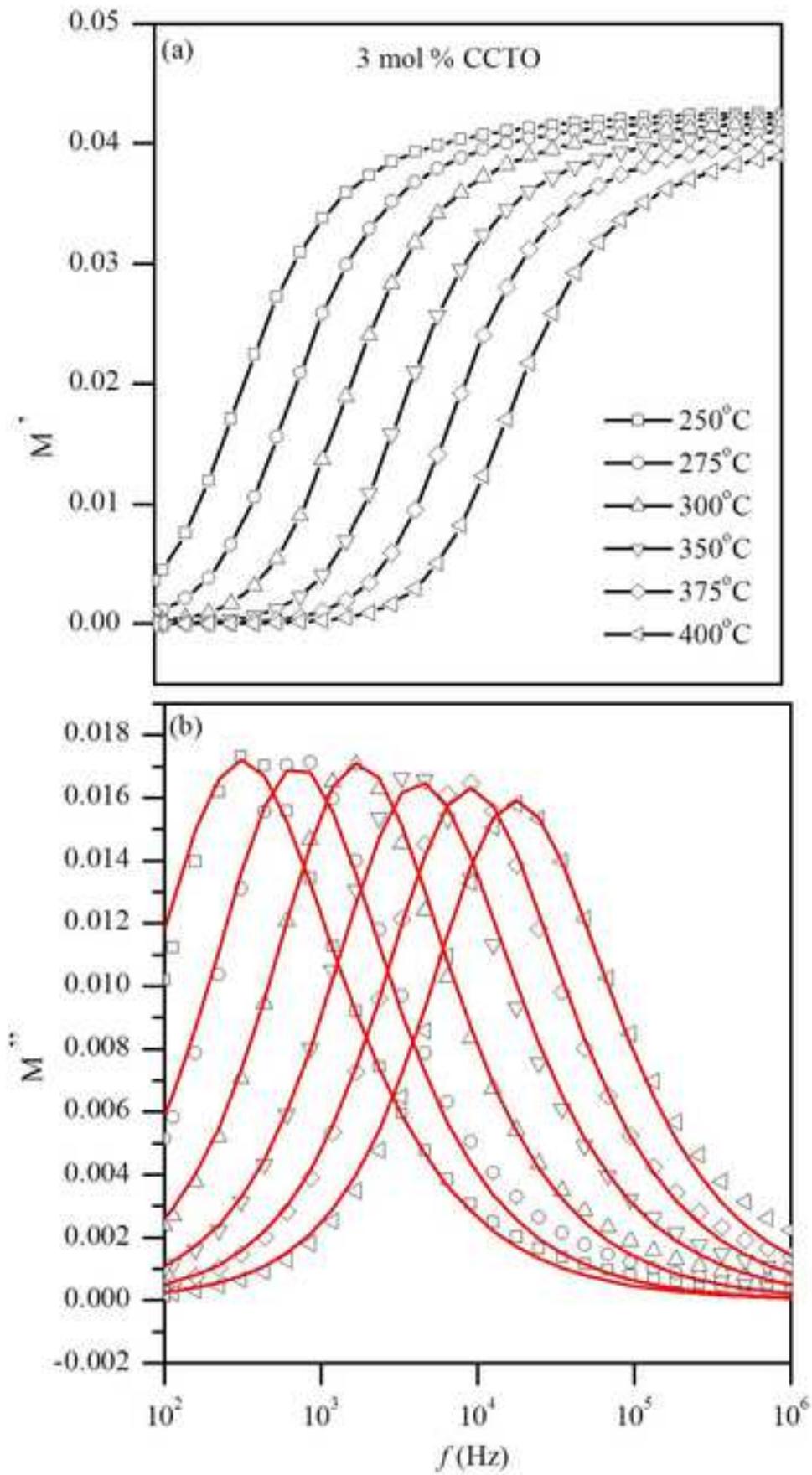

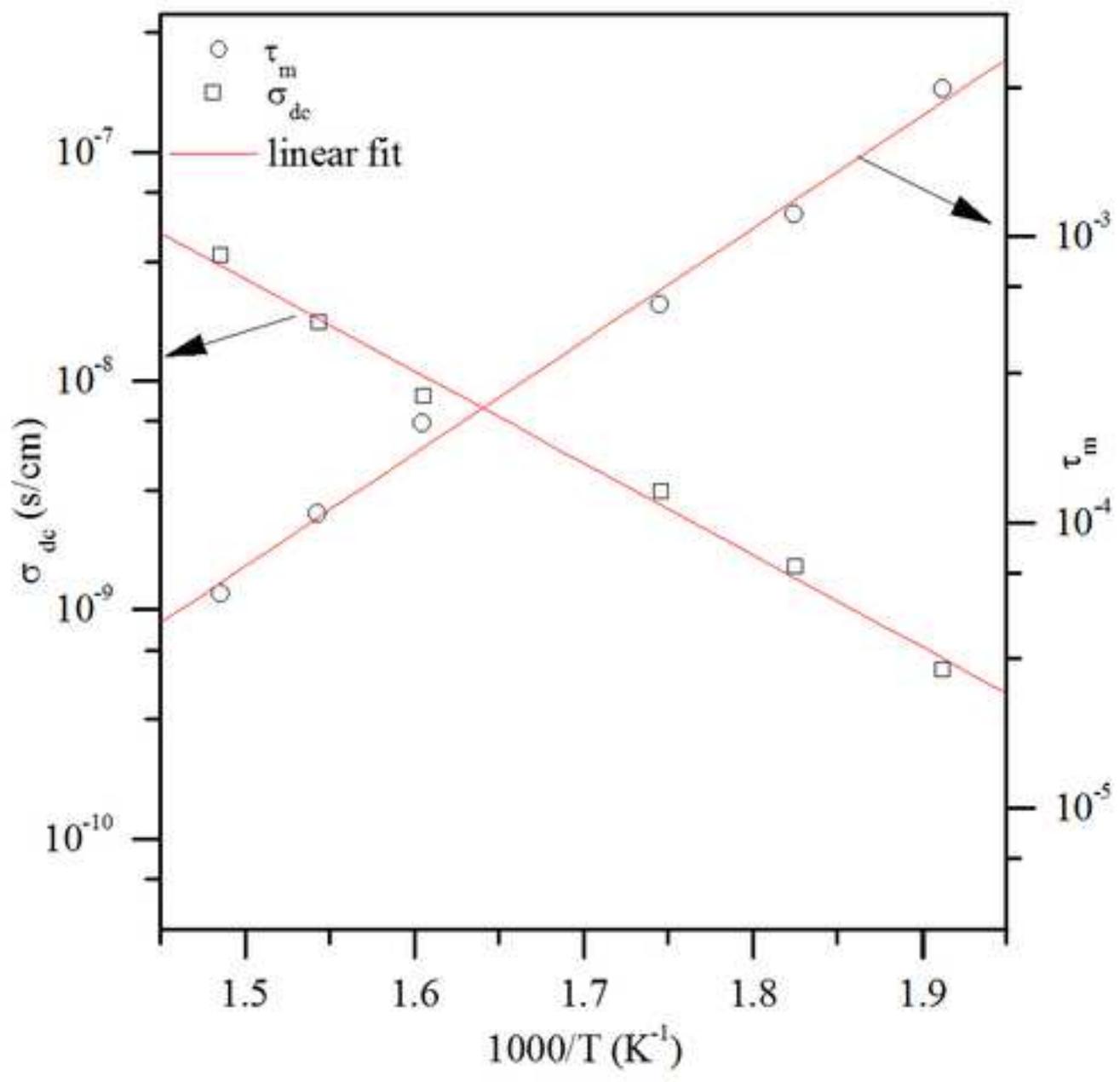